\documentclass[letterpaper, 10 pt, conference]{ieeeconf}  

\IEEEoverridecommandlockouts                              

\usepackage{xcolor}
\usepackage{amsmath}
\usepackage{amssymb}
\usepackage{mathtools}
\usepackage{dsfont}
\usepackage{comment}
\usepackage{flushend}

\DeclareMathOperator*{\argmin}{arg\min}
\usepackage{graphicx}
\usepackage{cite}
\usepackage{hyperref}

\newtheorem{remark}{Remark}

\title{\LARGE \bf
Data-Driven Multi-Modal Learning Model Predictive Control
}

\author{Fionna B. Kopp and Francesco Borrelli
\thanks{The authors are with the Department of Mechanical Engineering, University of California at Berkeley, Berkeley, CA 94701 USA}
}

\begin{document}

\maketitle
\thispagestyle{empty}
\pagestyle{empty}

\begin{abstract}
We present a Learning Model Predictive Controller (LMPC) for multi-modal systems performing iterative control tasks. Assuming availability of historical data, our goal is to design a data-driven control policy for the multi-modal system where the current mode is unknown. First, we propose a novel method to select local data for constructing affine time-varying (ATV) models of a multi-modal system in the context of LMPC. Then we present how to build a sampled safe set from multi-modal historical data.  We demonstrate the effectiveness of our method through simulation results of automated driving on a friction-varying track.

\end{abstract}

\section{Introduction}\label{sec:Intro}

Many methods exist to use stored data to design a control policy for iterative control problems including model-based and model-free approaches~\cite{rizk, bristow}. In this paper, we consider model-based methods which use data efficiently to find feasible trajectories and improve closed loop performance of iterative control tasks when the system dynamics are parameter-varying.
For example, it is well-known that road friction affects vehicle dynamics, particularly at the limits of handling, as it determines the minimum stopping distance as well as the attainable lateral forces and hence the achievable turning radius. Road friction can vary due to factors including weather, contaminant spills, or variation in road material which can be difficult to anticipate.
Our objective is to design a controller for automated driving based on a historical data set, which includes data from driving on different road frictions, to safely respond to abrupt changes in road friction.

One approach to parameter-varying systems in adaptive control is direct parameter estimation. This requires the classical parameter estimation/control pipeline. 
This paper instead explores the use of data-driven learning-based control ~\cite{rosolia2018learning, Hewing}. We focus on  iterative control tasks, and use Learning MPC (LMPC). LMPC is a reference-free iterative learning control method which uses historical data to construct terminal constraints and terminal cost functions~\cite{rosolia2018learning, rosolia2020racecar, rosoliaLMPCLinearSys}. LMPC uses stored state-input trajectories to build localized system models in real-time~\cite{rosolia2018learning, rosolia2020racecar, DeePC, LTVMPC}. 
Prior work relies on the assumption that the system dynamics are constant~\cite{rosolia2018learning, rosolia2020racecar, rosoliaLMPCLinearSys}. This allows for the entire data set of prior iterations to be representative of the system. We show that the same assumption on the data set and resulting approach is not able to handle rapidly changing conditions in real-time for a multi-modal system. 

In fact, modeling the system becomes a challenge in multi-modal systems where the dynamics can rapidly change, for example due to unknown and varying road friction. 
Assuming availability of multi-modal historical data, we propose a modified LMPC policy for the multi-modal system.

Similar attempts to data-driven learning-based approaches for system identification and control in multi-modal systems were proposed in~\cite{vaskov, EGP-MPC, NNMPC, GP_MM-MPC_HRI}. The authors in~\cite{vaskov} learn a Gaussian process (GP) of the friction curve to formulate a stochastic nonlinear model predictive controller for trajectory tracking under friction-varying conditions. In~\cite{EGP-MPC}, the authors propose using an ensemble GP from a library of GPs pre-trained on varying road conditions to model the system in a MPC framework for autonomous driving in varying surface conditions. The authors in~\cite{NNMPC} suggest using a neural network model to predict the vehicle dynamics in changing friction conditions within a MPC framework. Beyond autonomous driving, in~\cite{GP_MM-MPC_HRI} the authors propose building GP models for varying modes of human-robot interaction in a MPC control framework.

The contribution of this paper is threefold. First, we present an system identification strategy to learn local linear time-varying (LTV) models for control of a multi-modal system. Using the local LTV models, we present a method to construct a sampled safe set based on historical data. We use the LTV models and the safe set in a LMPC framework and demonstrate the effectiveness of our method through simulation results of automated driving on a friction-varying track.

\section{Problem Formulation}\label{sec:PD}

Consider the system described by the following dynamical model
\begin{equation}\label{eq:truesystem}
    x_{t+1} = f(x_t, u_t, \theta_t)
\end{equation}
subject to state and input constraints
\begin{equation}\label{eq:constraints}
    x_t \in \mathcal{X}, u_t \in \mathcal{U}, \quad \forall t \geq 0.
\end{equation}
Vectors $x_t \in \mathbb{R}^n$ and $u_t \in \mathbb{R}^m$ collect the states and inputs at time $t$. The parameter $\theta \in \mathbb{R}^p$ is time-varying and assumed to be unknown at runtime. Parameter $\theta$ is assumed to belong to parameter set $\Theta$,
\begin{equation}
    \theta_t \in \Theta, \quad \forall t \geq 0.
\end{equation}
At time step $t$, the nonlinear system~\eqref{eq:truesystem} can be linearized about $x_t$ and approximated as a quasi-Linear Parameter Varying (qLPV) system~\cite{NMPC_qLPV}
\begin{equation}
    x_{t+1} = A_t(\theta) x_t + B_t(\theta)u_t + g_t(\theta).
\end{equation}
When parameter set $\Theta$ is a discrete set, $\Theta = \{ \theta_1, \dots, \theta_n \}, \label{eq:parameterset}$ then system~\eqref{eq:truesystem} can be described as a hybrid system~\cite{SwitchLPV}, also referred to as a switched or multi-modal system. 
We use the term `multi-modal' in this paper.
We assume that we have a input/state data set $\mathcal{D}$ of the multi-modal system~\eqref{eq:truesystem} for multiple parameters $\theta \in \Theta$. 

The objective of this paper is to design a control policy for  system~\ref{eq:truesystem} which respects state and input constraints and minimizes an objective function by using local linear time-varying (LTV) approximations of $f(x, y, \theta)$ using the historical data $\mathcal{D}$.

\begin{remark} \label{remark:datalabels}
    Data in $\mathcal{D}$ does not need to be labeled as a function of the corresponding discrete parameter in $\Theta$. Even if the data is labeled by the corresponding parameter $\theta$, for instance when using a simulator to collect trajectories in different modes, the value of $\theta_t$ is unknown in real-time and thus $\mathcal{D}$ cannot be filtered according the true value of $\theta_t$.
\end{remark}
\begin{remark} \label{remark:datarichness}
    For the proposed method to be successful, we assume some level of richness in the data set. More precisely, we assume $\mathcal{D}$ contains data of the switched system~\ref{eq:truesystem} for multiple parameters $\theta \in \Theta$. It is reasonable to expect, but do not demonstrate it in this paper, that the success of the proposed method improves with the richness of the data set.
\end{remark}

A classical approach to this problem would consist of an offline system identification component to identify the dynamics $\hat{f}$ (since $f$ could be unknown or too complex for model-based design), a real-time parameter estimator to obtain the best estimate $\hat{\theta}_t$ of the real parameter $\theta_t$, and a state feedback controller. The literature on these topics is vast and beyond the scope of this paper~\cite{ferrari-trecate, bemporad, zhu}. This paper proposes a simpler data-driven architecture based on MPC and local learning of linear models. Similar to~\cite{rosolia2020racecar}, our approach makes use of learned local LTV models which are allowed to vary over the MPC prediction horizon. The LTV models are learned from local data which may vary along the prediction horizon, particularly in the multi-modal case, resulting in linearized models that vary over the prediction horizon. Further detail on the learned models is presented in section~\ref{sec:systemID}.

The goal can be formalized with the following constrained optimal control problem:
\begin{subequations}\label{eq:OCP}
    \begin{align}
        J_{0 \rightarrow \infty}^{*}(x_S) = \min_{u_0, u_1, \dots} & \sum_{k = 0}^T h(x_{k}, u_{k}) \label{subeq:OCP}\\
        \text{s.t.} \  & x_0 = x_S \label{subeq:OCPinit}\\
        & x_{k+1} = f(x_k, u_k, \theta_k), \: \forall k \geq 0 \label{subeq:OCPdyn} \\
        & x_{k} \in \mathcal{X}, u_{k} \in \mathcal{U}, \: \forall k \geq 0 \label{subeq:OCPconstr} \\
        & \theta_k \in \Theta , \: \forall k \geq 0 \label{subeq:OCPthetaconstr}\\
        & x_T \in \mathcal{T} \label{subeq:OCPTerminalConstr}
    \end{align}
\end{subequations}
where equations~\eqref{subeq:OCPinit} and~\eqref{subeq:OCPdyn} represent the initial condition and the system dynamics.  Equation~\eqref{subeq:OCPconstr} enforces the state and input constraints. Equation~\eqref{subeq:OCPthetaconstr} enforces parameter $\theta$ to be in discrete parameter set $\Theta$. Equation~\eqref{subeq:OCPTerminalConstr} enforces the final state $x_T$ to be in a non-empty target set $\mathcal{T} \subseteq \mathcal{X}$. The stage cost $h(x_k, u_k)$ in Eqn.~\eqref{subeq:OCP} is assumed to be continuous, jointly convex, and satisfy
\begin{align}
    h(x_F, 0) = 0 \: \text{and} \: h(x_t, u_t) \succ 0 \: \forall & x_t \in \mathbb{R}^n \setminus \{x_F\}, \nonumber \\
    & u_t \in \mathbb{R}^m \setminus \{0\}. \nonumber
\end{align}

In the constrained optimal control problem~\eqref{eq:OCP} both the parameter $\theta_k$ and the multi-modal dynamics $f$ are unknown at runtime. Even in the case where $\theta$ and $f$ are known, problem~\eqref{eq:OCP} is computationally demanding due to the large number of optimization variables. Instead, we aim to approximate a solution to problem~\eqref{eq:OCP} using a data-driven approach.
The objective of this paper is to design a control policy which uses historical data $\mathcal{D}$ of the multi-modal system~\eqref{eq:truesystem} to obtain an approximate solution to problem~\eqref{eq:OCP} where at time $t$ we only use $x_t$ and the historical dataset to learn the dynamics.

To do so, we propose the following strategy. We assume that through simulation or experiments historical data set $\mathcal{D}$ contains input/state data of system~\eqref{eq:truesystem} with different parameters $\theta_t \in \Theta$, where $\Theta$ is a discrete parameter set~\eqref{eq:parameterset}. We then propose a method to select local training data from $\mathcal{D}$, which represents the closest previously observed mode to the current system under unknown $\theta_k$. Using the selected local training data, we approximate the dynamics $f(x_t, u_t, \theta_t)$~\eqref{eq:truesystem} as a LTV model. Finally, we construct a control policy in real-time based on the data-driven LTV model of the true dynamics $f(x_t, u_t, \theta_t)$~\eqref{eq:truesystem}.

The proposed strategy fits naturally in the data-driven structure of learning model predictive control (LMPC) which approximates solutions to~\eqref{eq:OCP} for iterative control tasks~\cite{rosolia2018learning} in the case of uni-modal and known dynamics $f$. Hence, our work will make use of the LMPC structure, which is described in the following section.

This paper is organized as follows: in Section~\ref{sec:LMPC} we give an introduction to LMPC. Section~\ref{sec:MMstoreddata} describes our assumptions on the available historical data of the multi-modal system. Section~\ref{sec:kernelmethod} describes how local training data for the multi-modal system is selected from the historical data and how we estimate the switched dynamics from the selected subset of data. In Section~\ref{sec:MM-LMPC} we describe how to build a control policy for a multi-modal system using the LMPC structure. Finally, in Section~\ref{sec:results} we illustrate this approach through simulation results of autonomous driving on a friction-varying track.

\section{LMPC Introduction} \label{sec:LMPC}
In this section, we briefly introduce building LMPC policies for an iterative control task. LMPC proposes an approach to approximate the optimal control problem~\eqref{eq:OCP} by repeatedly solving a finite time optimal control problem (FTOCP) in a receding horizon fashion~\cite{rosolia2018learning}. More specifically, LMPC uses data from prior iterations to construct local convex terminal constraints and terminal cost function for a FTOCP, which is then solved in a receding horizon manner~\cite{rosolia2018learning}. 
Next we describe the components used to construct the LMPC policy that will be used in our proposed strategy. For more detail on LMPC refer to~\cite{rosolia2018learning, rosolia2020racecar, rosoliaLMPCLinearSys}.

\subsection{Stored Data}
Consider an iterative control task, we assume the closed-loop trajectories and associated input sequence of the $j$-th iteration are stored as follows,
\begin{align}
\begin{split} \label{eq:closedlooptrajs}
    \mathbf{u}^j = [u_0^j, u_1^j, \dots, u_t^j, \dots], \\
    \mathbf{x}^j = [x_0^j, x_1^j, \dots, x_t^j, \dots]
\end{split}
\end{align}
where $x_t^j$ and $u_t^j$ are the state and control input at time $t$ of iteration $j$. We assume that the stored closed-loop trajectories~\eqref{eq:closedlooptrajs} are feasible with respect to the state and input constraints~\eqref{subeq:OCPconstr}.

\subsection{Local Convex Safe Set} \label{subsec:LMPC_CSS}
Using a subset of the of the stored trajectories, the local convex safe set is constructed as in~\cite{rosolia2020racecar}. Specifically, the local convex safe set around state $x$ is the convex hull of the $M$-nearest neighbors to $x$ from selected iterations of the historical data.

For the $j$-th trajectory we define the set of time indices $[t_1^{j,*}, \dots, t_M^{j,*}]$ corresponding to the $M$-nearest neighbors of $x$,
\begin{align*}
    [t_1^{j,*}, \dots, t_M^{j,*}] = \argmin_{t_1, \dots, t_M} & \sum_{n = 1}^M \|x_{t_i}^j - x \|_D^2 \nonumber \\
    \text{s.t. } & t_i \neq t_k, \forall i \neq k \nonumber \\
    & t_i \in [0, \dots, T^j], \forall i \in [1, \dots, M]
\end{align*}
where $T^j$ is the time to complete iteration $j$ and matrix $D$ is a user-defined matrix to relatively weight the different state elements. The $M$-nearest neighbors to $x$ from a set of iterations $\{j_1, \dots, j_m \}$ are collected in the following matrix,
\begin{equation}
    D^{\{j_1, \dots, j_m\}}(x) = [x_{t_1^{j_1,*}}^{j_1}, \dots, x_{t_M^{j_1,*}}^{j_1}, \dots, x_{t_1^{j_m,*}}^{j_m}, \dots, x_{t_M^{j_m,*}}^{j_m}]
\end{equation}
which is used to construct the local convex safe set around $x$, 
\begin{align} \label{eq:LMPC_CSS}
    \mathcal{CL}^{\{j_1, \dots, j_m\}}(x) = \{\bar{x} \in \mathbb{R}^n \: : \: \exists &\mathbf{\lambda} \in \mathbb{R}^{Mm}, \mathbf{\lambda} \geq 0,  \mathds{1}\mathbf{\lambda} = 1, \nonumber \\
    & D^{\{j_1, \dots, j_m\}}(x)\mathbf{\lambda} = \bar{x} \}.
\end{align}
Then $\mathcal{CL}^{\{j_1, \dots, j_m\}}(x)$ is the convex hull of the $M$-nearest neighbors to $x$ from the iterations $\{j_1, \dots, j_m\}$.

\subsection{Local Convex Cost-to-Go}\label{subsec:LMPC_c2g}
As in~\cite{rosolia2020racecar}, we use the stored data to construct an approximation of the cost-to-go over the local convex safe set $\mathcal{CL}^{\{j_1, \dots, j_m\}}(x)$ around $x$. We define the local convex cost-to-go, $Q$-function, around $x$ as the convex combination of the costs associated with the stored trajectories used to construct the convex safe set,
\begin{align} \label{eq:LMPC_c2g}
    Q^{\{j_1, \dots, j_m\}}(x) = \min_{\mathbf{\lambda}} \: & \mathbf{J}^{\{j_1, \dots, j_m\}}(x) \mathbf{\lambda} \\
    \text{s.t. } & \mathbf{\lambda} \geq 0, \mathds{1} \mathbf{\lambda} = 1, \nonumber \\
    & D^{\{j_1, \dots, j_m\}}(x)\mathbf{\lambda} = \bar{x} \nonumber
\end{align}
where the vector
\begin{align}
    \nonumber
    \mathbf{J}^{\{j_1, \dots, j_m\}}(x) = [&J_{t_1^{j_1,*} \rightarrow T^{j_1}}^{j_1}(x_{t_1^{j_1,*}}^{j_1}), \dots, J_{t_M^{j_1,*} \rightarrow T^{j_1}}^{j_1}(x_{t_M^{j_1,*}}^{j_1}), \\
    & \dots, J_{t_1^{j_m,*} \rightarrow T^{j_m}}^{j_m}(x_{t_1^{j_m,*}}^{j_m}), \nonumber \\
    & \dots J_{t_M^{j_m,*} \rightarrow T^{j_m}}^{j_m}(x_{t_M^{j_m,*}}^{j_m})]. \nonumber
\end{align}
collects to cost-to-go associated with the $M$-nearest neighbors to $x$ from the selected iterations in set $\{j_1, \dots, j_m\}$. The cost-to-go $J_{t \rightarrow T^{j_i}}^{j_i}(x_t^{j_i}) = T^{j_i} - t$ represents to time to complete iteration $j_i$ from state $x_t^{j_1}$.

\subsection{LMPC Design} \label{subsec:LMPC_design}
At time $t$ of iteration $j$, LMPC solves the following finite time constrained optimal control problem~\cite{rosolia2018learning}:
\begin{subequations}\label{eq:LMPC_FTOCP}
    \begin{align}
        J_{t \rightarrow t+N}^{\text{LMPC}, j}(x_t^j) = 
        \min_{\mathbf{u}_t^j, \mathbf{\lambda}_t^j} & \left[\sum_{k=t}^{t+N-1} h(x_{k|t}, u_{k|t}) \nonumber \right.\\
        & \left. + \mathbf{J}^{\{j_1, \dots, j_m\}}(\bar{x}_{k+N})\mathbf{\lambda}_t^j \right] \\
        \text{s.t. } & x_{t|t} = x_t^j \label{subeq:LMPC_FTOCP_IC}\\
        & x_{k+1|t} = f(x_{k|t}, u_{k|t}), \nonumber\\
        &\forall k \in [t, \dots, t+N-1] \label{subeq:LMPC_FTOCP_sysdyn}\\
        & x_{k|t} \in \mathcal{X}, \: u_{k|t} \in \mathcal{U}, \nonumber \\
        & \forall k \in [t, \dots, t+N-1] \label{subeq:LMPC_FTOCP_constr} \\
        & \lambda_t^j \geq 0, \mathds{1}\lambda_t^j = 1, \nonumber \\
        & x_{t+N|t} \in D^{\{j_1, \dots, j_m\}}(\bar{x}_{k+N})\mathbf{\lambda}_t^j \label{subeq:LMPC_FTOCP_SS}
    \end{align}  
\end{subequations}
where $\mathbf{u}_t^j = [u_{t|t}^j, \dots, u_{t+N|t}^j]$ and $\bar{x}$ is the candidate solution which is chosen as the solution to~\eqref{eq:LMPC_FTOCP} from the prior time step. More precisely, the candidate solution $\bar{x}_{k+N} = x_{k+N|t-1}^*$. Equations~\eqref{subeq:LMPC_FTOCP_IC} and~\eqref{subeq:LMPC_FTOCP_sysdyn} represent the initial condition and system dynamics. Equation~\eqref{subeq:LMPC_FTOCP_constr} defines the state and input constraints. The constraint in~\eqref{subeq:LMPC_FTOCP_SS} forces the terminal state into the local convex safe set as described in Section~\ref{subsec:LMPC_CSS}.

Note the dynamics in Eqn.~\eqref{subeq:LMPC_FTOCP_sysdyn} are nonlinear. The nonlinear dynamics can be linearized around $x_t^j$ to obtain an affine time-varying (ATV) model of the system which allows the LMPC problem~\eqref{eq:LMPC_FTOCP} to be formulated as a quadratic program (QP) which can be solved efficiently. In a data-driven approach, the ATV models can be fully or partially learned from data as proposed in~\cite{rosolia2020racecar}. To learn the dynamics from data, we assume our historical data set $\mathcal{D}$ contains the previous closed loop trajectories~\eqref{eq:closedlooptrajs}. The learned ATV model is then determined by a linear regression of points $x_{k+1}^j$ on $x_{k}^j$ in $\mathcal{D}$. As in~\cite{rosolia2020racecar}, in order to select the locally relevant data to use for regression, the set $I^j(x)$ is defined as
\begin{align} \label{eq:NNindices}
    I^j(x) = \argmin_{\{k_1, j_1\}, \dots \{k_P, j_P\}} & \sum_{i=1}^P \| x - x_{k_i}^{j_i} \|_Q^2 \\
    \text{s.t. } & k_i \neq k_n, \forall j_i = j_n \nonumber \\
    & k_i \in \{1, 2, \dots\} \forall i \in \{1, \dots, P\} \nonumber \\
    & j_i \in \{1, \dots, j\}, \forall i \in \{1, \dots, P\}. \nonumber
\end{align}
The set $I^j(x)$ collects the indices associated with the $P$-nearest neighbors of $x$ from the stored data. The selected points are weighted by a chosen kernel function $K$, in our case we choose to use the Epanechnikov function~\cite{epanechnikov},
\begin{equation} \label{eq:kernel}
    K(u) =
    \begin{dcases}
        \frac{3}{4}(1-u^2), & \text{for } |u| < 1\\
        0 & \text{else}. \\
    \end{dcases}
\end{equation}
Equations~\eqref{eq:NNindices} and~\eqref{eq:kernel} are used to construct data-driven ATV models used for control in problem~\eqref{eq:LMPC_FTOCP}. At time $t$ of the $j$-th iteration, the ATV model takes the form as in~\cite{rosolia2020racecar}
\begin{equation} \label{eq:LMPC_ATVModel}
    x_{k+1|t}^j = A_{k|t}^j x_{k|t}^j + B_{k|t}^j u_{k|t}^j + g_{k|t}^j.
\end{equation}
Reformulating problem~\eqref{eq:LMPC_FTOCP} with the learned linear dynamics~\eqref{eq:LMPC_ATVModel}, we replace Eqn.~\eqref{subeq:LMPC_FTOCP_sysdyn} with~\eqref{eq:LMPC_ATVModel},
which can be efficiently solved by existing solvers. Let
\begin{align}
\begin{split}
\label{eq:optsoln}
    &\mathbf{u}_{t:t+N|t}^{*,j} = [u_{t|t}^{*,j}, \dots, u_{t+N-1|t}^{*,j}] \\
    &\mathbf{x}_{t:t+N|t}^{*,j} = [x_{t|t}^{*,j}, \dots, x_{t+N|t}^{*,j}]
\end{split}
\end{align}
denote the optimal solution of the problem at time $t$ of iteration $j$ with corresponding optimal cost $J_{t \rightarrow t+N}^{\text{LMPC}, j}(x_t^j)$. At time $t$ of the $j$-th iteration, the first element of $\mathbf{u}_{t:t+N|t}^{*,j}$ is applied to the system~\eqref{eq:truesystem}
\begin{align*}
    u_t^j = u_{t|t}^{*,j}.
\end{align*}
The FTOCP with the ATV dynamics is repeated at time $t+1$ based on the new state $x_{t+1|t+1} = x_{t+1}^j$, yielding a receding horizon strategy which approximates problem~\ref{eq:OCP}.

In prior work on LMPC~\cite{rosolia2018learning, rosoliaLMPCLinearSys, rosolia2020racecar}, there was no notion of a switched or multi-modal system. Rather, the prior works assume constant system dynamics~\cite{rosolia2018learning, rosoliaLMPCLinearSys, rosolia2020racecar}. The contribution of this work is to explain how to construct the local convex safe set, cost-to-go function, and local ATV models to build a LMPC control policy for multi-modal systems when the mode in unknown in real-time.

The remainder of this paper is organized as follows. Section~\ref{sec:MMstoreddata} briefly describes the structure and assumptions on the stored data of the multi-modal system. Section~\ref{sec:kernelmethod} outlines how the nearest neighbors and kernel function in Eqns.~\eqref{eq:NNindices} and~\eqref{eq:kernel} are modified for the multi-modal system and how we use the identified local training data to construct data-driven local ATV models~\eqref{eq:ATVModel} for control. Finally, in Section~\ref{sec:MM-LMPC} we describe how to construct the local convex safe set~\eqref{eq:LMPC_CSS} and cost-to-go function~\eqref{eq:LMPC_c2g} to build a multi-modal LMPC control policy. We demonstrate our proposed approach through simulation results of autonomous driving on a friction-varying track in Section~\ref{sec:results}.

\section{Multi-Modal System Identification}\label{sec:systemID}
\subsection{Stored Data}
\label{sec:MMstoreddata}
In this section, we outline our assumptions on the stored data of the multi-modal system~\eqref{eq:truesystem}. Let the entire data set be denoted $\mathcal{D}$ while $\mathcal{D}_\theta$ refers to data of the system for particular parameter $\theta$. $\mathcal{D}_\theta$ contains state-input trajectories and the local ATV model approximations as
\begin{equation} \label{eq:Dtheta}
    \mathcal{D}_\theta = \left \{ \bigcup_{i \in R_\theta^j} \bigcup_{t=0}^N (u_t^i, x_t^i, A_t^i, B_t^i, g_t^i) \right \}
\end{equation}
where $R_\theta^j$ is the set of indices $r$ associated with successful iterations under the mode defined by $\theta$. For $r \leq j$,
\begin{equation}\label{eq:Rindexset}
    R_\theta^j = \{ r \in [0, j] : \theta_t^r = \theta, \forall \, x_t^r \in \mathcal{X}, t \geq 0\}.
\end{equation}

Next we show how $A_t^j, B_t^j$ and $g_t^j$ are constructed using the state-input trajectories $x_t^i, u_t^i$ from prior iterations $i < j$. The linearized dynamics, $A_t^j, B_t^j$ and $g_t^j$, included in the historical data $\mathcal{D}$ are used to select iterations to build the local convex safe set~\eqref{eq:LMPC_CSS} for the LMPC policy. This is further detailed in Section~\ref{sec:MM-LMPC}.

Going forward, we will remove the dependence of $\mathcal{D}$ on parameter $\theta$. In fact,  as highlighted in Remark~\ref{remark:datalabels}, the proposed algorithm does not require a labeled data set, i.e. the knowledge of the true value of $\theta$ for a given state-input pair in the data set. 

\begin{remark}
    A system simulator can be used to construct an initial nonempty data set $\mathcal{D}^0 = \{ \mathcal{D}_{\theta_1}, \mathcal{D}_{\theta_2}, \dots \}$ by explicitly setting parameter values. Though the parameter labels of the simulated data set are known at the time of construction, the data set is unlabeled going forward.
\end{remark}

\subsection{Local Data Identification and Weighting} \label{sec:kernelmethod}
In this section, we describe how to select local training data closest to the current observed mode in real-time. More specifically, we describe the needed modifications to Eqn.~\eqref{eq:NNindices} and to the kernel function~\eqref{eq:kernel} to select and weight the local training data for the multi-modal system.

At time step $k$ with unknown parameter $\theta_k$, the unknown dynamics $f(x_k, u_k, \theta_k)$ are learned from the stored trajectories and approximated by a local linearized model,
\begin{equation}\label{eq:linearsys}
    x_{k+1} = A_k x_k + B_k u_k + g_k.
\end{equation}
The linearized model in~\eqref{eq:linearsys} is given by a linear regression of the points $x_{k+1}^j$ on $x_k^j$ from the stored trajectories.

At time step $k$, let us define the tuple 
\begin{equation} \label{eq:zbar}
    \bar{z}_k = (x_{k|t}, u_{k|t-1}^*, x_{k+1|t-1}^*).
\end{equation}
Note that the input $u_{k|t-1}^*$ and next state $x_{k+1|t-1}^*$ components of $\bar{z}_k$ are from the optimal solution~\eqref{eq:optsoln} at the prior time step $t-1$ because the model regression is performed prior to the solving the problem at time step $t$. Let us also represent the state-input trajectories in $\mathcal{D}$ as tuples $z_k^j = (x_k^j, u_k^j, x_{k+1}^j)$. At time step $k$ with tuple $\bar{z}_k$ the data set $\mathcal{D}$ is queried to find the $P$-nearest neighbors to $\bar{z}_k$. We find the set of indices for the $P$-nearest neighbors to $\bar{z}_k$ using Eqn.~\eqref{eq:NNindices} modified as follows with $\bar{z}_k$
\begin{align} \label{eq:NNindiceszbar}
    I^j(\bar{z}_k) = \argmin_{\{k_1, j_1\}, \dots \{k_P, j_P\}} & \sum_{i=1}^P \| \bar{z}_k - z_{k_i}^{j_i} \|_Q^2 \\
    \text{s.t. } & k_i \neq k_n, \forall j_i = j_n \nonumber \\
    & k_i \in \{1, 2, \dots\} \forall i \in \{1, \dots, P\} \nonumber \\
    & j_i \in \{1, \dots, j\}, \forall i \in \{1, \dots, P\}. \nonumber
\end{align}
\begin{remark}
    As the data set $\mathcal{D}$ is unlabeled with respect to parameter $\theta$, the set of $P$-nearest neighbors to $\bar{z}_k$ can be from multiple previously observed modes. In this way, the $P$-nearest neighbors identify nearest one-step dynamic tuples. In the case when the observed mode is not directly in the prior data set, this allows for the $P$-nearest neighbors to still be selected regardless of their respective mode and represent the closest one-step dynamic behavior.
\end{remark}

Using the identified $P$-nearest neighbors and the Epanechnikov function~\eqref{eq:kernel}, the weighted least squares problem is defined as:
\begin{equation} \label{eq:leastsquares}
    \Gamma(\bar{z}) = \argmin_{\Gamma} \sum_{\{k,j\} \in I(\bar{z})} K \left( \frac{\|\bar{z}-z_k^j\|_{Q}^2}{\eta} \right) y^j_k(\Gamma)
\end{equation}
where $Q$ is a user defined weighting matrix and $\eta \in \mathbb{R}_+$ is a bandwidth.
Note the Epanechnikov kernel function in Eqn.~\eqref{eq:leastsquares} is modified to weight differences on $\bar{z}$ rather than state alone for the multi-modal case. The optimizer of~\eqref{eq:leastsquares}, is used to compute the local linearized model~\eqref{eq:linearsys}.

This approach is similar to the method from~\cite{rosolia2020racecar, LTVMPC}. The key difference here is the use of the predicted input $u_{k|t-1}^*$ and the predicted next state $x_{k+1|t-1}^*$ in equations~\eqref{eq:NNindiceszbar} and~\eqref{eq:leastsquares} to select and weight the data points used for model regression of the multi-modal system.  

\begin{remark} \label{remark:kernel}
    In prior work~\cite{rosolia2020racecar}, the system dynamics $x_{t+1} = f(x_t, u_t)$ are assumed to be constant for all iterations, which implies the collected data is under constant system dynamics. All prior iterations are representative of the current dynamics and simply the nearest neighbors to the state $x$ or state-input pair $(x,u)$ can be used to identify local data for model regression. However, for the multi-modal system the data set $\mathcal{D}$ contains data of the system under multiple modes. In order to obtain the closest corresponding mode in the data to the current mode, our approach implicitly searches for prior data with similar dynamic matrices, $A, B$ and $g$. This is performed by searching for nearest data in the form of the tuples $z$ which implicitly contain information about the dynamic behavior of the system and hence the mode.
\end{remark}

Next we describe how to construct the ATV model~\eqref{eq:ATVModel} for the multi-modal system, which closely follows~\cite{rosolia2020racecar}. At time $t$ of iteration $j$, the optimal solution~\eqref{eq:optsoln} from time $t-1$ is used to define the candidate solution $\bar{\mathbf{x}}_t^j = [ \bar{x}_{t|t}^j, \dots, \bar{x}_{t+N|t}^j]$. Using the candidate solution, the following ATV model is constructed
\begin{equation} \label{eq:ATVModel}
    x_{k+1|t}^j = A_{k|t}^j x_{k|t}^j + B_{k|t}^j u_{k|t}^j + g_{k|t}^j
\end{equation}
where $A_{k|t}^j, B_{k|t}^j,$ and $g_{k|t}^j$ are obtained by linearizing the dynamics $f(x_k, u_k, \theta_k)$ by solving the least squares problem~\eqref{eq:leastsquares} at $\bar{z}_k^j$. For more detail, we refer the reader to~\cite{rosolia2020racecar}. 
\begin{remark}
    Because $\bar{z}_k^j$ is time-varying over the prediction horizon as defined in Eqn.~\eqref{eq:zbar}, the ATV model in Eqn.~\eqref{eq:ATVModel} is not only time varying with each iteration of the FTOCP, but also over the LMPC prediction horizon. 
\end{remark}

\section{Multi-Modal LMPC (MM-LMPC) Design} \label{sec:MM-LMPC}
In this section, we describe how to use the identified system model from~\ref{sec:systemID} and data set $\mathcal{D}$ to construct the local convex safe set~\eqref{eq:LMPC_CSS} and terminal cost function $Q$~\eqref{eq:LMPC_c2g} for multi-modal LMPC (MM-LMPC).

In LMPC, the terminal safe set is constructed from the state trajectories of prior iterations as described in Section~\ref{subsec:LMPC_CSS}~\cite{rosolia2018learning, rosolia2020racecar, rosoliaLMPCLinearSys}.
Prior works assume the dynamics are constant, so the local safe set can be constructed from the immediately preceding or best performing iterations~\cite{rosolia2018learning, rosolia2020racecar, rosoliaLMPCLinearSys}. However in the multi-modal system, our data set $\mathcal{D}$ contains data from multiple modes, necessitating a method to select which iterations to use for the local convex safe set~\eqref{eq:LMPC_CSS} and terminal cost function~\eqref{eq:LMPC_c2g}.

Ideally the safe set represents the set of states from which there exists a series of control actions to complete the task, ie. the safe set is an under approximation of the maximal stabilizable set~\cite{rosolia2018learning}. In other works, there exist guarantees on this fact~\cite{rosolia2018learning, rosoliaLMPCLinearSys}; however, this relies on the assumption that the dynamics are constant. Here, in the multi-modal case we want to find a safe set that is reachable in the current mode and constructed from trajectories under a similar mode.

\begin{remark}\label{remark:MM-LMPC_SS}
    To give intuition, if the safe set is constructed from prior iterations with a similar mode, the safe set is more likely to represent states from which there exists a series of control actions to complete the task in the current mode. In order to determine which iterations are similar, we compare the estimated state trajectory under our current ATV model to the trajectories predicted by the stored ATV models. Similar to Remark~\ref{remark:kernel}, comparing trajectories implicitly captures and compares the dynamics of the system and so is used to implicitly determine similarity between modes.
\end{remark}

At time step t of the $j$-th iteration, the system's state is $x_t^j$ and the dynamics are linearized by the ATV model from Eqn.~\eqref{eq:ATVModel}. At time $t$, the optimal solution~\eqref{eq:optsoln} at time $t-1$ is used to construct the candidate control inputs over the prediction horizon,
\begin{equation}\label{eq:candidate_inputs}
    \begin{split}
        \bar{\mathbf{u}}_t^j & = [ \bar{u}_{t|t}^j, \dots, \bar{u}_{t+N-1|t}^j ] \\
        & = [u_{t|t-1}^{*,j}, \dots, u_{t+N-1|t-1}^{*,j}].
    \end{split}
\end{equation}
With our ATV model~\eqref{eq:ATVModel} and the candidate inputs~\eqref{eq:candidate_inputs}, we construct $\hat{\mathbf{x}}_t^j = [\hat{x}_{t|t}^j, \dots \hat{x}_{t+N|t}^j]$,
\begin{equation}
    \hat{x}_{k+1|t}^j = A_{k|t}^j \hat{x}_{k|t}^j + B_{k|t}^j \bar{u}_{k|t} + g_{k|t}, \: \forall k \in [t, \dots, t+N-1]
\end{equation}
where $\hat{x}_{t|t}^j = x_t^j$. Namely, $\hat{\mathbf{x}}_t^j$ collects the predicted state trajectory from current state $x_t^j$ under the current ATV model.

To find the similar prior iterations, we want to compare the current predicted trajectory $\hat{\mathbf{x}}_t^j$ to the state trajectories predicted by applying the same candidate control inputs to the stored ATV models in $\mathcal{D}$. To do so, we similarly define predicted state trajectories under the ATV models in $\mathcal{D}$ for all iterations $j_i < j$.

First, we find the closest state in each prior iteration to the current state $x_t^j$. Let us define the closest state from prior iteration $j_i<j$ as $x^{j_i}_{t_{j_i}^*}$ where $t_{j_i}^*$ is the time index
\begin{align*}
    t_{j_i}^* = \argmin_{t_{j_i}} \|x_t^j - x_{t_{j_i}}^{j_i}\|_1.
\end{align*}

For all $j_i < j$ we compute the predicted state trajectory $\hat{\mathbf{x}}^{j_i} = [\hat{x}_{t_{j_i}^*}^{j_i}, \dots, \hat{x}_{t_{j_i}^*+N}^{j_i}]$,
\begin{equation}
    \hat{x}_{k+1}^{j_i} = A_k^{j_i} \hat{x}_k^{j_i} + B_k \bar{u}_{k} +g_k^{j_i}, \forall k \in \{ t_i^*, \dots, t_i^*+N-1 \}
\end{equation}
where we initialize with the current state $\hat{x}_{t_{j_i}^*}^{j_i} = x_t^j$ and 
\begin{align*}
    \bar{u}_{k_i} = \bar{u}_{k|t}, \: \forall &k_i \in [t_i^*, \dots, t_i^*+N-1], \nonumber\\
    &k \in [t, \dots, t+N-1].
\end{align*}
For each prior iteration $j_i < j$, $\hat{\mathbf{x}}^{j_i}$ collects the predicted state trajectory under the model at iteration $j_i$ due to the candidate control inputs from iteration $j$.

The predicted trajectory $\hat{\mathbf{x}}_t^j$ under the current ATV model is compared to the predicted trajectories $\hat{\mathbf{x}}^{j_i}$ to find the $N_{SS}$ iterations with the most similar predicted trajectories. More precisely,
\begin{align} \label{eq:MMLMPCiters}
    L(\hat{\mathbf{x}}_t^j) = \argmin_{j_i < j} \sum_{i = 1}^{N_{SS}} & \|\hat{x}_{t|t}^j - \hat{x}_{t_{j_i}^*}^{j_i} \|_1 + \dots \nonumber \\
    &+ \|\hat{x}_{t+N|t}^j - \hat{x}_{t_{j_i}^*+N}^{j_i} \|_1,
\end{align}
such that $L(\hat{\mathbf{x}}_t^j)$ collects the set of indices associated with the $N_{SS}$ prior iterations with most similar predicted trajectories from current state $x_t^j$ due to control inputs $\bar{\mathbf{u}}_t^j$.

If the difference in predicted trajectories,
\begin{equation}\label{eq:modeldiff}
    \|\hat{x}_{t|t}^j - \hat{x}_{t_{j_i}^*}^{j_i} \|_1 + \dots + \|\hat{x}_{t+N|t}^j - \hat{x}_{t_{j_i}^*+N}^{j_i} \|_1 < \delta, \: \forall j_i \in L(\hat{\mathbf{x}}_t^j)
\end{equation}
where $\delta \in \mathbb{R}_+$ is a user chosen bandwidth, then the local convex safe set~\eqref{eq:LMPC_CSS} is constructed using stored state trajectories from iterations $L(\hat{\mathbf{x}}_t^j)$. In other words, we calculate $\mathcal{CL}^{L(\hat{\mathbf{x}}_t^j)}$ using Eqn.~\eqref{eq:LMPC_CSS}. Otherwise if the condition in Eqn.~\eqref{eq:modeldiff} does not hold, a MPC safety controller is used to reference track a reference trajectory until the predictions are within bandwidth $\delta$ and Eqn.~\eqref{eq:modeldiff} holds. The bandwidth parameter $\delta$ and the MPC safety controller are user-determined.

\begin{remark}
    The use of the safety controller should be minimized by tuning $\delta$, to rely primarily on the prior data while maintaining state constraints. Unless the data set $\mathcal{D}$ contains data of the system under the safety controller, then the more the braking controller is used to reference track a slower velocity reference, the resulting state evolution is more likely to deviate from previous iterations in $\mathcal{D}$. With more deviation from modes in $\mathcal{D}$, there may be limited or no data within the threshold $\eta$ for model regression in Section~\ref{sec:systemID}.
\end{remark}

As described in Section~\ref{subsec:LMPC_c2g} the convex terminal cost function is defined over the local convex safe set. In the multi-modal case, the identified iterations $L(\hat{\mathbf{x}}_t^j)$ from Eqn.~\eqref{eq:MMLMPCiters} are used to determine the cost-to-go function~\eqref{eq:LMPC_c2g}.

Using the identified indices $L(\hat{\mathbf{x}}_t^j)$ to construct the local convex safe set and local cost-to-go $Q$-function for the multi-modal case, we propose solving the FTOCP~\eqref{eq:LMPC_FTOCP} with the dynamics~\eqref{subeq:LMPC_FTOCP_sysdyn} replaced by~\eqref{eq:ATVModel} as described in Section~\ref{sec:kernelmethod} and the set of indices for the local convex safe set and cost-to-go replaced with $L(\hat{\mathbf{x}}_t^j)$ from Eqn.~\eqref{eq:MMLMPCiters} to obtain a MM-LMPC control policy.
        
\section{Results} \label{sec:results}
\begin{figure*} [!ht]
    \centering
    \includegraphics[scale=0.7]{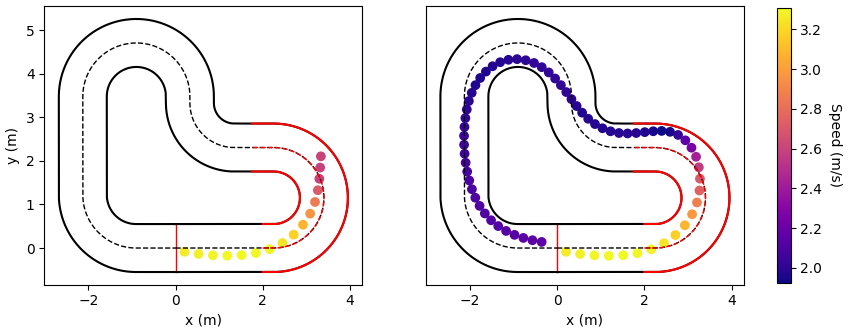}
    \vspace{-0.3cm}
    \caption{Comparison of the closed loop trajectories of the $j$-th iteration with low (red) and high (black) road friction. The trajectory from the prior approach~\cite{rosolia2020racecar, LTVMPC} to system identification and construction of ATV models is shown on the left and our proposed approach on the right.}
    \vspace{-0.5cm}
    \label{fig:model_estimation_closed_loop}
\end{figure*}
\begin{figure} [htbp]
    \centering
    \includegraphics[scale=0.6]{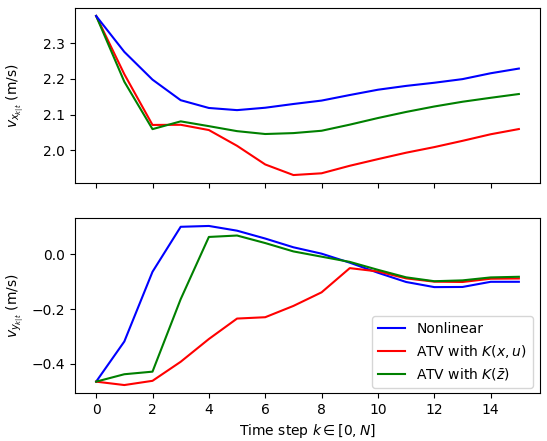}
    \vspace{-0.2cm}
    \caption{Comparison of the open loop trajectories of the velocity states over the prediction horizon in the multi-modal system.}
    \vspace{-0.5cm}
    \label{fig:model_est_open_loop}
\end{figure}
\begin{figure} [htbp]
    \centering
    \includegraphics[scale=0.6]{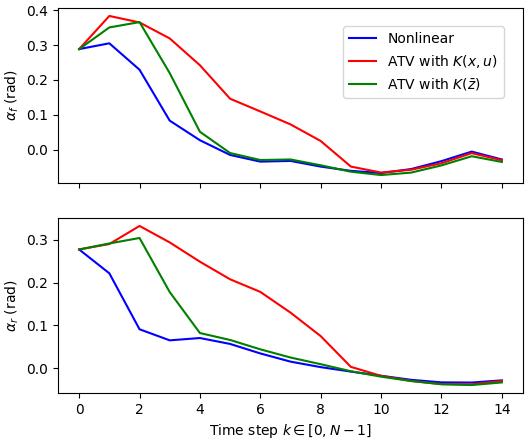}
    \vspace{-0.2cm}
    \caption{Comparison of the predicted slip angles over the prediction horizon in the multi-modal system.}
    \vspace{-0.5cm}
    \label{fig:model_est_open_loop_slip_angles}
\end{figure}
\begin{figure*} [!t]
    \centering
    \includegraphics[scale=0.7]{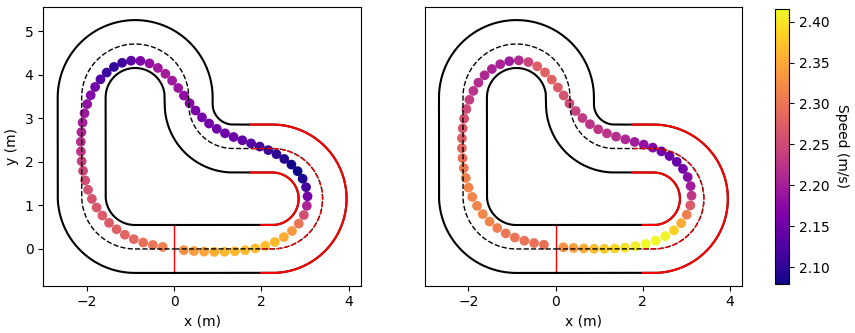}
    \vspace{-0.3cm}
    \caption{Comparison of the converged laps of LMPC~\cite{rosolia2020racecar} initialized on the track with both low (red) and high (black) road friction (left) and our proposed approach after re-converging after iteration $j$ when the multi-modal system switches to the combined low and high friction track (right).}
    \vspace{-0.5cm}
    \label{fig:converged_lap_LMPC_MMLMPC}
\end{figure*}
\begin{figure} [htbp]
    \centering
    \includegraphics[scale=0.6]{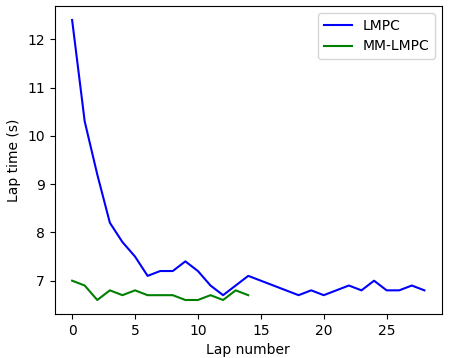}
    \vspace{-0.2cm}
    \caption{Lap time of LMPC initialized on the constant-mode combined-friction track compared with MM-LMPC after iteration $j$ when the multi-modal system switches to the combined friction track.}
    \vspace{-0.5cm}
    \label{fig:LMPC_MMLMPC_lap_times}
\end{figure}

The proposed approach is implemented in simulation for autonomous driving on a friction-varying track. For the context of this example, we work with the planar dynamic bicycle model with the following state and input vectors~\cite{kong2015bicycle}
$$
x = [v_x, v_y, \omega_z, e_\psi, s, e_y]^\top \text{ and } u = [a, \delta]^\top
$$
where $v_x, v_y, \omega_z$ are the vehicle's longitudinal velocity, lateral velocity, and the yaw rate. The position of the vehicle is represented in the Frenet frame with respect to a parametric path where $s, e_y, e_\psi$ are the distance traveled along the path, the lateral distance from the path, and the heading angle between the path tangent and the vehicle. The inputs are longitudinal acceleration $a$ and steering angle $\delta$. In this example, the considered system parameter $\theta$ is one-dimensional and equivalent to road friction $\mu$.

We assume that the dynamics may be be split into $f_1(x_k, u_k)$ which is known and independent of parameter $\theta$ and $f_2(x_k, u_k, \theta_k)$ which is unknown and parameter-dependent,
$f(x_k, u_k, \theta_k) = f_1(x_k, u_k) + f_2(x_k, u_k, \theta_k).$
With respect to the dynamic bicycle model, the kinematics are well-known and independent of road friction $\mu$~\cite{rosolia2020racecar, kong2015bicycle}. Hence, the kinematics will be represented by $f_1$. The dynamic states, $v_x, v_y, \omega_z$, depend on physical parameters including road friction $\mu$~\cite{rosolia2020racecar}. Here $f_2$ represents the dynamic states. For other multi-modal systems, $f_1$ and $f_2$ may be chosen accordingly.

Let the dynamic states be collected in the set $l = \{ v_x, v_y, \omega_z\}$. Then we write the tuple from Eqn.~\eqref{eq:zbar} with respect to the set $l$ as $\bar{z}_k^l = (x_{k|t}, u_{k|t-1}^*, x_{k+1|t-1}^{l,*})$.
For the least squares regression~\eqref{eq:leastsquares} here the vector $\Gamma \in \mathbb{R}^5$ and
\begin{equation} \nonumber
    y_k^{j,l}(\Gamma) = \|x_{k+1}^{j,l} - \Gamma[v^j_{x_{k}}, v^j_{y_{k}}, \omega^j_{z_{k}}, a^j_k, 1]^\top \|.
\end{equation}
The optimizer of the least squares regression~\eqref{eq:leastsquares} is used to linearize $f_2$ around $x$ as follows:
\begin{align} \label{eq:linearizeddynamics}\nonumber
    \begin{bmatrix}
        v_{x_{k+1}} \\
        v_{y_{k+1}} \\
        \omega_{z_{k+1}} \\
    \end{bmatrix} = &
    \begin{bmatrix}
        \Gamma_{1:3}^{v_x}(\bar{z}) \\
        \Gamma_{1:3}^{v_y}(\bar{z}) \\
        \Gamma_{1:3}^{\omega_z}(\bar{z})
    \end{bmatrix}
    \begin{bmatrix}
        v_{x_k} \\
        v_{y_k} \\
        \omega_{z_k}
    \end{bmatrix} \\
    & + \begin{bmatrix}
        \Gamma_4^{v_x}(\bar{z}) & 0 \\
        0 & \Gamma_4^{v_y}(\bar{z}) \\
        0 & \Gamma_4^{\omega_z}(\bar{z})
    \end{bmatrix}
    \begin{bmatrix}
        a_k \\ \delta_k
    \end{bmatrix} + \begin{bmatrix}
        \Gamma_5^{v_x}(\bar{z})\\
        \Gamma_5^{v_y}(\bar{z})\\
        \Gamma_5^{\omega_x}(\bar{z})
    \end{bmatrix}
\end{align}
where $\Gamma_i^l(\bar{z})$ is the $i$th element of $\Gamma^l(\bar{z})$. At every time step $k$, the learned local linear model~\eqref{eq:linearizeddynamics} of unknown $f_2$ is combined with the kinematic model $f_1$ linearized about the state $x$ to obtain the full linearized system dynamics in~\eqref{eq:linearsys}. The ATV model~\eqref{eq:ATVModel} described in Section~\ref{sec:kernelmethod} is obtained by linearizing $f_1$ around $\bar{x}_{k|t}^j$ and by evaluating~\eqref{eq:linearizeddynamics} at $\bar{z}_{k}^j$.

Next we compare our method to compute local ATV models to the approach from~\cite{rosolia2020racecar, LTVMPC} on a multi-modal system with unknown friction and to the true nonlinear dynamics with known friction $\mu$.
 
For this example, the initial data set $\mathcal{D}^0$ is comprised of two modes on a L-shaped track: a high friction track $\mu = 0.9$ and a low friction track $\mu = 0.6$.
In this example, let iteration $j-1$ be on the high-friction track.
Upon iteration $j$, the track changes such that $25 \%$ of the track beginning at $2m$ is low friction $\mu = 0.6$ while the rest of the track remains high friction $\mu = 0.9$.

Figure~\ref{fig:model_estimation_closed_loop} compares the closed loop trajectories of the $j$-th iteration of the multi-modal system using the prior method to generate ATV models~\cite{rosolia2020racecar, LTVMPC} (left) and our proposed approach (right). The low friction region is indicated by the red track outline. With the prior approach, the vehicle fails to maintain state constraints, falling off the track. With our proposed approach, the vehicle maintains state constraints and completes iteration $j$ under the new mode.

The open loop trajectories of the velocities over the prediction horizon are illustrated in Figure~\ref{fig:model_est_open_loop}. This provides insight as to why our proposed approach results in improved closed loop behavior for multi-modal systems. The addition of $x_{k+1|t-1}^*$ in $\bar{z}$ provides information about the system's response to input $u_{k|t-1}^*$. As mentioned in Remarks~\ref{remark:kernel} and~\ref{remark:MM-LMPC_SS}, by comparing similarity between state propagation, rather than $(x,u)$ or $x$ alone, our proposed method implicitly compares the dynamic behavior and hence mode. Consequently, our approach is more likely to identify the subset of historical data that represents the observed mode, resulting in a better approximation of the true model.

Similarly, the predicted slip angles in open loop over the prediction horizon are shown in Figure~\ref{fig:model_est_open_loop_slip_angles}. More accurate estimation of the slip angles and velocities in the open loop is imperative for maintaining state constraints of a multi-modal system in closed loop, especially in a constrained environment such as that of a track.

Our proposed approach to system identification
enables the control policy to maintain state and input constraints while completing iteration $j$ in the new mode. With the completion of iteration $j$, data under the new mode is added to $\mathcal{D}$. 
Recall that the initial data set consists of two constant modes, while at iteration $j$ the system exhibits a case of spatially-varying friction, not directly in the initial data set.
By adding data from iteration $j$, the data set is augmented both with respect to the spatially-varying mode as well as the fact that the state trajectories seen in low friction region are different than the low friction data in the initial data set because the high-to-low transition is not captured in the original data set.

Our proposed approach to MM-LMPC allows for the data set to be augmented as described, enabling the control policy to quickly learn and converge in the new mode.
Figure~\ref{fig:converged_lap_LMPC_MMLMPC} compares the converged iteration of previously proposed LMPC policy~\cite{rosolia2020racecar} initialized on the multi-friction track (left) to the converged iteration of the MM-LMPC policy where for the $j$-th iteration and after the track consists of both high and low friction sections. With the proposed approach, the MM-LMPC policy re-converges in $15$ laps after the change at iteration $j$. The LMPC policy proposed in~\cite{rosolia2020racecar} initialized on multi-friction track converged in $28$ laps after PID initialization laps. This comparison can be thought of as the worst-case baseline--if the system had to fully reinitialize and relearn after iteration $j$ it would take $28$ laps. Figure~\ref{fig:LMPC_MMLMPC_lap_times} illustrates the lap times, demonstrating that MM-LMPC can re-converge to the same performance in the multi-modal case.

\section{Conclusion}
We presented a multi-modal Learning Model Predictive Controller (MM-LMPC) for multi-modal systems performing iterative control tasks. The proposed framework uses historical data of the system to estimate the dynamics of the observed mode, and construct safe sets and approximated terminal cost functions, and consequently design a data-driven control policy for multi-modal systems. We demonstrated the effectiveness of the proposed strategy on simulated autonomous driving on a friction-varying track. The results demonstrated improved system and mode identification and closed loop-performance for the multi-modal system.

\bibliographystyle{ieeetr}
\bibliography{root}

\begin{thebibliography}{10}

\bibitem{rizk}
H.~Rizk, A.~Chaibet, and A.~Kribèche, ``Model-based control and model-free control techniques for autonomous vehicles: A technical survey,'' {\em Applied Sciences}, vol.~13, 2023.

\bibitem{bristow}
D.~Bristow, M.~Tharayil, and A.~Alleyne, ``A survey of iterative learning control,'' {\em IEEE Control Systems Magazine}, vol.~26, no.~3, pp.~96--114, 2006.

\bibitem{rosolia2018learning}
U.~Rosolia and F.~Borrelli, ``Learning model predictive control for iterative tasks. a data-driven control framework,'' {\em IEEE Transactions on Automatic Control}, vol.~63, pp.~1883--1896, July 2018.

\bibitem{Hewing}
L.~Hewing, K.~P. Wabersich, M.~Menner, and M.~N. Zeilinger, ``Learning-based model predictive control: Toward safe learning in control,'' {\em Annual Review of Control, Robotics, and Autonomous Systems}, vol.~3, pp.~269--296, 2020.

\bibitem{rosolia2020racecar}
U.~Rosolia and F.~Borrelli, ``Learning how to autonomously race a car: A predictive control approach,'' {\em IEEE Transactions on Control Systems Technology}, vol.~28, pp.~2713--2719, November 2020.

\bibitem{rosoliaLMPCLinearSys}
U.~Rosolia and F.~Borrelli, ``Learning model predictive control for iterative tasks: A computationally efficient approach for linear systems,'' {\em IFAC-PapersOnLine}, vol.~50, no.~1, pp.~3142--3147, 2017.

\bibitem{DeePC}
J.~Coulson, J.~Lygeros, and F.~Dörfler, ``Data-enabled predictive control: In the shallows of the deepc,'' in {\em 2019 18th European Control Conference (ECC)}, pp.~307--312, 2019.

\bibitem{LTVMPC}
D.~Papadimitriou, U.~Rosolia, and F.~Borrelli, ``Control of unknown nonlinear systems with linear time-varying mpc,'' in {\em 2020 59th IEEE Conference on Decision and Control (CDC)}, pp.~2258--2263, 2020.

\bibitem{vaskov}
S.~Vaskov, R.~Quirynen, M.~Menner, and K.~Berntorp, ``Friction-adaptive stochastic predictive control for trajectory tracking of autonomous vehicles,'' in {\em 2022 American Control Conference (ACC)}, pp.~1970--1975, 2022.

\bibitem{EGP-MPC}
T.~Nagy, A.~Amine, T.~X. Nghiem, U.~Rosolia, Z.~Zang, and R.~Mangharam, ``Ensemble gaussian processes for adaptive autonomous driving on multi-friction surfaces,'' {\em IFAC-PapersOnLine}, vol.~56, no.~2, pp.~494--500, 2023.

\bibitem{NNMPC}
N.~A. Spielberg, M.~Brown, and J.~C. Gerdes, ``Neural network model predictive motion control applied to automated driving with unknown friction,'' {\em IEEE Transactions on Control Systems Technology}, vol.~30, no.~5, pp.~1934--1945, 2022.

\bibitem{GP_MM-MPC_HRI}
K.~Haninger, C.~Hegeler, and L.~Peternel, ``Model predictive control with gaussian processes for flexible multi-modal physical human robot interaction,'' in {\em 2022 International Conference on Robotics and Automation (ICRA)}, pp.~6948--6955, 2022.

\bibitem{NMPC_qLPV}
P.~S.~G. Cisneros, S.~Voss, and H.~Werner, ``Efficient nonlinear model predictive control via quasi-{LPV} representation,'' in {\em 2016 IEEE 55th Conference on Decision and Control (CDC)}, pp.~3216--3221, 2016.

\bibitem{SwitchLPV}
H.~Atoui, O.~Sename, V.~Milanes, and J.-J. Martinez-Molina, ``Toward switching/interpolating {LPV} control: A review,'' {\em Annual Reviews in Control}, vol.~54, pp.~49--67, 2022.

\bibitem{ferrari-trecate}
G.~Ferrari-Trecate, D.~Mignone, and M.~Morari, ``Moving horizon estimation for hybrid systems,'' {\em IEEE Transactions on Automatic Control}, vol.~47, no.~10, pp.~1663--1676, 2002.

\bibitem{bemporad}
A.~Bemporad, D.~Mignone, and M.~Morari, ``Moving horizon estimation for hybrid systems and fault detection,'' in {\em Proceedings of the 1999 American Control Conference}, pp.~2471--2475, 1999.

\bibitem{zhu}
F.~Zhu and P.~J. Antsaklis, ``Optimal control of hybrid switched systems: A brief survey,'' {\em Discrete Event Dynamic Systems: Thoery and Applications}, vol.~25, pp.~345--364, 2014.

\bibitem{epanechnikov}
V.~A. Epanechnikov, ``Non-parametric estimation of a multivariate probability denisty,'' {\em Theory of Probability \& Its Applications}, vol.~14, no.~1, pp.~153--158, 1969.

\bibitem{kong2015bicycle}
J.~Kong, M.~Pfeiffer, G.~Schildbach, and F.~Borrelli, ``Kinematic and dynamic vehicle models for autonomous driving control design,'' in {\em IEEE Intelligent Vehicles Symposium (IV)}, pp.~1094--1099, 2015.

\end{thebibliography}

\end{document}